# Encapsulation of DNA by Cationic Diblock Copolymer Vesicles

A. V. Korobko, W. Jesse, and J. R. C. van der Maarel*

*Leiden Institute of Chemistry, Leiden University, P. O. Box 9502,
2300 RA Leiden, the Netherlands*

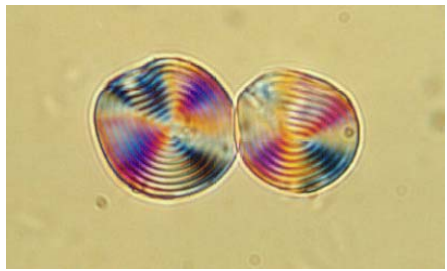

————————————————

*Corresponding author. Voice: +31 715274543, Fax: +31 71 5274603,
E-mail: j.maarel@chem.leidenuniv.nl

Encapsulation of dsDNA fragments (contour length 54 nm) by the cationic diblock copolymer poly(butadiene-*b*-N-methyl 4-vinyl pyridinium) [PBd-*b*-P4VPQ] has been studied with phase contrast, polarized light, and fluorescence microscopy, as well as scanning electron microscopy. Encapsulation was achieved with a single emulsion technique. For this purpose, an aqueous DNA solution is emulsified in an organic solvent (toluene) and stabilized by the amphiphilic diblock copolymer. The PBd block forms an interfacial brush, whereas the cationic P4VPQ block complexes with DNA. A subsequent change of the quality of the organic solvent results in a collapse of the PBd brush and the formation of a capsule. Inside the capsules, the DNA is compacted as shown by the appearance of birefringent textures under crossed polarizers and the increase in fluorescence intensity of labeled DNA. The capsules can also be dispersed in aqueous medium to form vesicles, provided they are stabilized with an osmotic agent (polyethylene glycol) in the external phase. It is shown that the DNA is released from the vesicles once the osmotic pressure drops below $10^5$ N/m$^2$ or if the ionic strength of the supporting medium exceeds 0.1 M. The method has also proven to be efficient to encapsulate pUC18 plasmid in sub-micron sized vesicles and the general applicability of the method has been demonstrated by the preparation of the charge inverse system: cationic poly(ethylene imine) encapsulated by the anionic diblock poly(styrene-*b*-acrylic acid).

# Introduction

Self-assembly is clearly a necessary tool to realize practical nanoscale structures. These nanostructures will probably involve membrane vesicles either as the nanostructures themselves, or as templates for more complex structures.[1] Because of the large molecular weight, compared to lipids, most polymer membranes are hyper thick, and can thereby achieve far greater stability than any natural lipid membrane.[2] Accordingly, although vesicles can also be composed of surfactants or lipids, polymeric vesicles are preferred. The present contribution aims at the fundamental design, control, and structural characterization of a new class of such complex, self-assembled structure: cationic copolymer vesicles loaded with DNA. These vesicles may serve as an experimental model system for diverse applications such as non-viral gene delivery and packaging of DNA in, *e.g.*, condensates, bacteriophages, and viruses.[3,4,5,6]

Micron and nano-size polymeric vesicles and capsules have considerable potential in industrial, medical, and pharmaceutical applications, because of their ability to take up and carry a reagent through an otherwise hostile medium. The usual preparation procedure involves the adsorption of alternating layers of oppositely charged polyelectrolytes on a colloidal particle.[7,8] In the next step, the colloidal core is destroyed either by UV irradiation or immersion in a strong acid or base. The resulting empty capsule can subsequently be loaded with a drug. A drawback of this method is that the molecular weight of the reagent is restricted by the pore size in the capsule membrane, because the reagent is usually introduced *after* the preparation of the capsule. This restriction makes it difficult to encapsulate rather large macromolecules such as clone vector DNA, although recently progress has been made by precipitation of spermidine condensed DNA onto the surface of template micro-particles.[9] There is a clear need for high enough compaction of DNA, efficiency of encapsulation, and control of the structure and properties of the protective shell.

Our encapsulation experiments are inspired by DNA-polycation complexes (polyplexes).[10] DNA forms complexes with cationic polymers, which can be used for targeting DNA into cells.[11,12] Polyplexes are formed when the DNA fuses with a cationic polymer and equal numbers of positive (from the DNA) and negative (from the polymer) small counterions are released into the bulk. This entropic gain due to counterion release is thought to be the driving force for the formation of the complex.[13] The cationic polymer can differ in chemical composition and architecture of the backbone, *e.g.*, linear, branched, or diblock copolymer. As a result, a wide range of condensing agents can be synthesized and evaluated in order to determine the most efficient system. A common feature of the previously investigated systems is the *micelle-like* structure of the complex coacervate.[14] The micelles contain a domain with DNA and the neutralizing polymer, possibly surrounded by a coronal layer composed of the neutral attachment of the copolymer.

It is our contention that the ability of ionic diblocks to form self-assembled *vesicular* structures, together with the counterion release mechanism as a driving force for the complex formation, offer new opportunities for the encapsulation of charged (bio)polymers. Although we will demonstrate our encapsulation procedure for DNA, we argue that the method is more general. Our method is based on a water-in-oil emulsion as template, but, contrary to a similar approach reported in the literature,[15] the encapsulated material is part of the membrane. We start with the preparation of an emulsion of aqueous droplets dispersed in an immiscible organic solvent and stabilized by a polyelectrolyte diblock copolymer. The droplets can contain any water-soluble macromolecule with opposite net charge to form a polyelectrolyte bilayer with the ionic block at the water/copolymer interface. The organic solvent should be a good solvent for the hydrophobic attachment. If the ionic block is sufficiently small, the good solvent quality ensures that the copolymer can be dissolved in the organic phase prior to mixing. The emulsion is then prepared by mixing the aqueous and organic phases in the right proportion. To a certain extent, the size of the droplets can be controlled by the emulsification procedure, *e.g.*, by sonication or micro-filtration. Long time stability of the emulsion can be achieved by the appropriate choice of the copolymer concentration and molecular weight of the hydrophobic attachment. There is no restriction to the molecular weight of the material in the aqueous phase, because at this stage of the preparation the capsule has not been formed.

For the formation of the capsule, the good organic solvent is replaced by a non-solvent, which results in a collapse of the hydrophobic chain part. The collapsed polymer layer now forms a shell around the droplet, as is illustrated in Figure 1. If the non-solvent is miscible with water, the material in the aqueous phase gets compacted because water is extracted from the droplet during the encapsulation process. The volatile organic solvent can subsequently be evaporated to form capsules in air. To produce vesicles, the capsules can then be transferred into an aqueous supporting medium. It has proven to be necessary to immerse the capsules in polyethylene glycol (PEG) solution rather than pure water, because otherwise, the membrane becomes unstable and the encapsulated material is released. The stability of the vesicles is determined by the balance of the elastic force of the membrane itself and the osmotic forces exerted by the encapsulated material and the supporting medium, respectively.

Apart from stabilizing the emulsion, the copolymer is also the building block of the polymer shell in the final stage of the preparation procedure. Accordingly, the stability and permeability of the membrane depend on the properties of the hydrophobic attachment (*e.g.*, glass temperature, molecular weight, *etc.*) as well as the properties of the polyelectrolyte bilayer formed by the ionic block and the encapsulated material at the inner side of the interfacial layer. Release of the encapsulated material, which is of great importance in application studies, can be affected by two different, but related mechanisms. The first one is an imbalance in the osmotic and elastic stretching forces acting on the membrane, whereas the second mechanism involves a change in the permeability of the membrane itself. An example of the first mechanism is the release of the encapsulated material once the osmotic pressure exerted by the supporting medium drops below a certain critical value. As an example of the second mechanism, we will see below that the permeability of the membrane can be controlled by an increase in ionic strength through screening of the electrostatic interactions in the polyelectrolyte bilayer.

The organization of this paper is as follows. First, we describe the step-wise production of the vesicles with the single emulsion technique. This process is followed with polarized light and scanning electron microscopy, as well as fluorescence microscopy of labeled DNA. Then we move on to the determination of the compaction efficiency and the DNA density profile inside the emulsion droplets, capsules in air, and vesicles in aqueous medium. The stability of the vesicles against osmotic pressure and ionic strength is investigated by controlled release experiments using fluorescence labeled DNA. We demonstrate that there are no restrictions to the molecular weight of the DNA by the encapsulation of clone vector, pUC18 plasmid (2686 base pairs). We will also show that the applicability of the method is general by the preparation of the 'charge inverse' system: cationic homopolymer encapsulated by an anionic diblock copolymer.

# Experimental Section

**Chemicals and Solutions.** DNA was obtained by micrococcal nuclease digestion of calf thymus chromatin.[16] After precipitation in cold 2-propanol, the DNA pellet was dried under reduced pressure at room temperature. The DNA was brought to the salt free sodium form by dissolving it in a 50 mM NaCl, 24 mM EDTA buffer and extensive dialysis against water (purified by a Millipore system with conductivity less than $10^{-6}$ $\Omega^{-1}\text{cm}^{-1}$). To avoid denaturation, care was taken that the DNA nucleotide concentration did not drop below 3 mM nucleotides/l. The differential molecular weight distribution was monitored by size exclusion chromatography (SEC) with light scattering detection.[17] The advantage of the isolation procedure is that it yields a large quantity of mononucleosomal DNA, but a typical batch contains approximately 25 % lower and higher molecular weight material.

Further SEC fractionation resulted in a relatively monodisperse eluent fraction with an average molecular weight $M_w$ = 104000 (158 base-pairs, contour length $L$ = 54 nm) with polydispersity $M_w/M_n$ = 1.14. The hypochromic effect at 260 nm confirmed the integrity of the double helix and the ratio of the optical absorbencies $A_{260}/A_{280}$ = 1.83 indicates that the material is essentially free of protein.[18] The material was freeze-dried and the residual water content was determined by IR spectroscopy. Stock solutions were prepared by dissolving freeze dried Na-DNA in $H_2O$. The DNA concentrations are 36, 18, 9, 4.5, and 2.3 g/l. The concentrations are determined by weight, using the water content in the freeze-dried material and the Na-DNA partial molar volume 165 $cm^3$/mol. As fluorescence DNA probe, 4',6-diamidino-2-phenylindole dichloride (DAPI) was added to the stock solutions with a concentration 58 μg of DAPI per gram of DNA. pUC18 plasmid was isolated from *Escherichia coli* bacteria, purified, and characterized as described in previous work.[19] The plasmid concentration in the stock solution is 1.4 g/l.

Poly(butadiene-*b*-N-methyl-4-vinyl-pyridinium iodide) [PBd-*b*-P4VPQI] diblock copolymer was purchased from Polymer Source Inc., Dorval, Canada. According to the manufacturer, the number-average molecular weights $M_n$ of the PBd and P4VPQI blocks are 120000 and 28200 g/mol, which correspond with a degree of polymerization DP = 2220 and 115, respectively. The molecular weight polydispersity $M_w/M_n$ ratio of the copolymer is 1.05. Solutions were prepared by dissolving the copolymer in toluene. The copolymer concentrations are determined by weight and have the values 4, 2, 1, 0.5, 0.1, and 0.01 g/l. Polyethyleneglycol (PEG) with molecular weight 3000 g/mol was purchased from Merck. Six osmotic stress solutions are prepared by dissolving the PEG in purified water. The PEG weight concentrations are 43, 21, 10, 5, 1, and 0.5 wt %. For the preparation of the charge inverse system, cationic poly(ethylene imine) (PEI) with molecular weight 35000 g/mol was purchased from Fluka. PEI was encapsulated with the anionic poly(styrene-*b*-acrylic acid) (PS-*b*-PA, Polymer Source Inc). The diblock copolymer is 'crew cut' with molecular weights 11000 and 1240 g/mol of the PS and PA blocks, respectively. The concentrations of the PEI in water and PS-*b*-PA in toluene stock solutions are 11.3 and 1.1 g/l.

**Imaging.** For light microscopy, a droplet of the emulsion was deposited on a microscope slide and sealed with a cover slip. DNA capsules were deposited on a microscope slide by the procedure as described below. The capsules were either directly observed in air or they were first immersed in PEG solution and sealed with a cover slip. Polarized light microscopy was done with a Leica DMR microscope with 10×, 63×, and 100× (oil immersion) objectives at ambient temperature. The magnification was calibrated with the help of a ruler. Images were collected with a Ricoh 35 mm photo camera. Phase contrast and fluorescence imaging were done with an Olympus BX - 60 microscope equipped with a 100 W mercury lamp and a UV filter set (U-MWU/ narrow band cube; excitation at 330-385 nm and with an emission filter at 420 nm). The exposure time of the DAPI fluorescence label was controlled by a UV light shutter. Images were collected with a CCD camera and analyzed with the public domain software Object-Image 2.06.[20] The radial fluorescence intensity of the emulsion droplets, capsules, or vesicles was measured and the background was subtracted. For scanning electron microscopy, capsules were deposited on a microscope slide and dried at 40 °C and 0.75 atm for one week. The specimens were subsequently sputter coated with gold under vacuum and imaged with a JEOL SEM 6400 microscope.

## Results and Discussion

**Production of the Vesicles.** We will first demonstrate that our procedure can be used to encapsulate DNA. The first step involves the preparation of an emulsion of aqueous DNA droplets dispersed in toluene and stabilized with copolymer. For this purpose, two solutions are prepared: one solution is made of the cationic diblock poly(butadiene-*b*-N-methyl-4-vinylpyridinium iodide) [PBd-*b*-P4VPQI] in toluene and the second one is an aqueous solution of 150 base-pair DNA fragments without added low molecular weight salt. The size of the polybutadiene attachment is sufficiently large to ensure a good solubility in toluene, despite the presence of the cationic block. To achieve long time stability of the emulsion, it proved to be necessary to dissolve 4 gram of copolymer per liter of toluene. For most experiments, the DNA concentration in the initial aqueous phase was 36 g/l. The aqueous DNA solution is then mixed with the copolymer in toluene solution in volume ratio 1:9, respectively, and stirred for several hours to form an emulsion. Due to the amphiphilic behavior of the copolymer, the hydrophobic polybutadiene attachment prevents the droplets from coalescence and the emulsion was observed to be stable over months.

Figure 2 displays an optical micrograph of an emulsion of DNA/water droplets immersed in toluene. The size of the droplets is on the order of tens of microns, but here no special efforts were done for size fractionation and/or to prepare droplets of smaller size. The DNA concentration inside the droplets amounts 36 g/l, which is well below the critical concentration pertaining to the transition to the cholesteric, liquid crystalline phase (100-220 g/l, depending on ionic strength).[21] With the help of polarized light microscopy, it was observed that the droplets are not birefringent and, hence, the DNA is indeed not liquid-crystalline. Fluorescence imaging with DAPI labeled DNA shows that the droplets contain DNA and that there is no significant amount of DNA present within the supporting toluene phase (see Figure 5).

The DNA molecules form a polyelectrolyte bilayer with the cationic block at the copolymer/water interface and contribute to the stability of the membrane. To gauge the importance of the polyelectrolyte bilayer, some experiments were done in which the toluene was evaporated from the emulsion. If the droplets do not contain DNA, *i.e.* if they are prepared with pure water, this process results in the formation of a copolymer film. In the presence of DNA however, the droplets remain spherical and capsules are formed (these capsules are covered by a copolymer film, results not shown). It was observed, however, that for the lowest 2.3 g/l DNA concentration the larger droplets collapse, whereas the integrity of the smaller ones is preserved. These results support the view that the electrostatic interaction between DNA and the cationic block controls the stability and permeability of the membrane. Indeed, as we will see below with fluorescence microscopy, screening of this interaction by the addition of low molecular weight salt results in the release of DNA.

In the second step of the preparation procedure, the emulsion is transferred into ethyl acetate, which is a non-solvent for polybutadiene. Figure 3a displays a polarized light micrograph of the emulsion during solvent exchange by capillary suction of ethyl acetate between slide and cover slip. The micrograph shows the coexistence of non-birefringent droplets and birefringent capsules. After completion of the solvent exchange, all droplets have become birefringent and the textures are reminiscent of those observed for liquid crystalline DNA.[22,23] Apart from the solvent quality induced collapse of the hydrophobic attachment, during the solvent exchange water is extracted from the capsule and the DNA is compacted in an orderly fashion imposed by the emulsion template. The compaction is facilitated by the good miscibility of water and ethyl acetate. Occasionally, we have observed the typical fingerprint-like textures for cholesteric DNA (as in Figure 3d), but in most cases the textures were quite irregular.

As shown by polarized light, scanning electron, and fluorescence microscopy in Figures 3b, 4, and 5b, respectively, the integrity of the capsules is preserved after evaporation of the volatile ethyl acetate. In particular, there is no change in polarized light microscopy textures over an extended period in time. The textures show clear cholesteric fringes, if the dry capsules are taken up in toluene again. This observation suggests that the rigidity of the interfacial layer prevents uniform alignment of the liquid crystal. As we will see below with fluorescence microscopy, the DNA concentration inside the dry capsules is indeed in the range of the one pertaining to the macroscopic cholesteric phase. The scanning electron micrographs show that the surface of the capsules is homogeneous and that there is a broad distribution in capsule size with average and minimum diameter 16 and 3 μm, respectively. Notice that the size is determined by the emulsification procedure. As will be shown below, sub-micron size capsules (with clone vector DNA) can be prepared if the emulsion is sonicated prior to the solvent quality induced collapse of the hydrophobic attachment. At room temperature, polybutadiene is well above the glass temperature and liquid-like, which explains the tendency of the dry capsules to become stuck together once the solvent is evaporated. However, this does not affect the long-term stability of the capsules and their ability to re-disperse in aqueous medium.

In the final stage of the preparation, vesicles are produced by transferring the capsules into an aqueous medium. However, the vesicles are unstable in pure water and it is necessary to use an osmotic agent in the external phase. Figures 3c and 3d display polarized light micrographs of the vesicles suspended in 43 wt % aqueous PEG solution. The micrographs show two types of liquid-crystalline textures in coexistence. All vesicles show a Maltese cross and blue and yellow interference colors generated by a full-wave retardation plate, which suggests that the DNA molecules are oriented perpendicular to the radius. Around 5 percent of the vesicles exhibit a cholesteric molecular arrangement (Figure 3d). In the radial direction away from the center, they show a periodicity with an alternation of dark and light stripes corresponding to half the cholesteric pitch. The pitch is 2 μm, which is the same value as the one reported for the macroscopic cholesteric phase.[22,23] The remaining vesicles do not show a periodicity in the radial direction, so there is no twist in orientation order with increasing distance away from the center (Figure 3c). The latter texture does not show, however, the characteristic fan-like shapes as reported for the macroscopic, high density, hexagonal phase.[22,23] The absence of the fan-like shapes might be related to the mesoscopic dimension and the geometric frustration effect of the spherical interface. The fact that we observe two different textures in coexistence indicates that the DNA concentration is in the range of the critical boundaries pertaining to the first order phase transition from the cholesteric to the hexagonal phase (280-410 g/l, depending on ionic strength).[24] This will be confirmed below with the help of fluorescence microscopy.

**Density Distribution and Compaction Factors.** In order to investigate the density distribution of the encapsulated material, we have done fluorescence microscopy with DAPI labeled DNA. Figure 5 displays the fluorescence micrographs of the emulsion droplets in toluene, dry capsules in air, and vesicles suspended in 43 wt % aqueous PEG solution. The radial dependencies of the corresponding, azimuthally averaged intensities are displayed in Figure 6. Notice that the radial coordinate has been scaled by the maximum radius $R$ and the intensities have been divided by the corresponding intensities measured at the center of the object. As can seen in Figure 6, the scaled intensities collapse to a single master curve, which shows that the intensity profiles are similar for DNA confined in an emulsion droplet, a capsule in air, or a vesicle in aqueous medium. If the density is uniform, the azimuthally averaged profile of the 2D projected image of a spherical object with maximum radius $R$ takes the form

$$I(r) = \alpha R \sqrt{1 - (r/R)^2} \tag{1}$$

where $r$ is the distance away from the center and $\alpha$ denotes a constant which is proportional to the DNA concentration. As can be seen in Figure 6, the radial intensities satisfy eq (1), which shows that the DNA distribution is indeed uniform. The deviations observed for $r/R$ > 1 are related to the optical resolution of the microscope.

For each object (*i.e.*, droplet, capsule, or vesicle in toluene, air, and aqueous medium, respectively), the radius $R$ and the fluorescence intensity at the center $I(r = 0)$ was determined from a fit of eq (1) to the radial fluorescence intensity profile. The results are displayed in Figure 7, for a population in radii between, say, 2 and 15 µm. For a uniformly filled spherical object and if the DNA concentration does not depend on the size, the fluorescence intensity at the center of the 2D image is linear in the radius

$$I(r = 0) = \alpha R \tag{2}$$

No systematic deviation from a linear dependence of $I(r = 0)$ versus $R$ in Figure 7 is observed. In particular, for the capsules in air and vesicles in aqueous medium the fluorescence intensities show a rather strong and random scatter. This scatter is not related to uncertainty in the measurements, but to a variation in DNA concentration between capsules or vesicles of comparable size (the average concentration is size independent). As already shown by the occurrence of birefringent textures, the concentration increases and, hence, the DNA gets progressively compacted from the emulsion, through the capsules in air, to the vesicles in aqueous medium. The concentrations can be derived from the slopes $\alpha$, because $\alpha$ is proportional to the DNA concentration and the DNA concentration inside the emulsion droplets is known. Results of the fit of eq (2) to the data in Figure 7 are collected in Table 1. From the ratios of the slopes with respect to the one pertaining to the emulsion (compaction factors), we obtain 240±40 and 350±50 g/l for the DNA concentration inside the capsules in air and vesicles in aqueous medium, respectively. Notice that our encapsulation procedure has resulted in a 10-fold compaction of DNA.

It is interesting to compare the DNA concentrations with the critical boundaries in the bulk phase diagram of DNA with the same molecular weight. The phase diagram shows two first-order phase transitions. Depending on ionic strength, the isotropic-cholesteric phase transition occurs between 100 and 220 g/l and the transition from the cholesteric to the hexagonal phase takes place in the region 280-410 g/l.[21,24] The DNA concentration inside the capsules in air, as obtained from fluorescence microscopy, is indeed in the cholesteric regime. However, the average DNA concentration inside the vesicles in aqueous medium is close to the critical boundaries pertaining to the first-order transition to the hexagonal phase. This explains the coexistence of vesicles exhibiting different polarized light microscopy textures: the structure inside around 95 percent of the vesicles is hexagonal and the remaining fraction is cholesteric. It also explains the observed scatter in fluorescence intensity; the difference in concentrations in the coexisting hexagonal and cholesteric bulk phases is in the range 5 to 10 percent.[24]

The fluorescence microscopy experiments support the view already obtained with polarized light microscopy. If the emulsion droplet is immersed in ethyl acetate, the hydrophobic attachment collapses and forms a membrane. During this process and after evaporation of the volatile organic solvent, water is extracted from the capsule until the osmotic pressure exerted by the DNA equilibrates the surface tension of the polymeric shell. The compaction, however, continues when the dry capsules are taken up in the aqueous PEG solution. Now, the osmotic pressure exerted by the supporting medium should be incorporated in the force balance. As a result, the vesicle further shrinks with a concomitant increase in DNA concentration and, hence, internal osmotic pressure until mechanical equilibrium is reached. As we will see below, a certain minimum external osmotic pressure is needed to balance the osmotic pressure exerted by the encapsulated DNA and to prevent release of the DNA into the aqueous medium. We will now further explore the stability of the vesicles against osmotic stress and ionic strength.

**Stability of the Vesicles.** Control of the stability of the vesicles and release of the encapsulated material are of great importance in application studies. The stability of the vesicles is achieved by the balance of forces acting on the membrane. This force balance involves the osmotic pressure exerted by the encapsulated DNA, the surface tension of the membrane itself, and the pressure exerted by the osmotic agent in the supporting medium. At least two different, but related release mechanisms can be envisioned. The first one is an imbalance in the osmotic and elastic stretching forces acting on the membrane. If the osmotic pressure exerted by the supporting medium decreases, the vesicle is expected to swell with a concomitant decrease in internal pressure and increase in membrane tension until the force balance is reestablished. However, if the external pressure drops below a certain critical value, the elasticity of the membrane might not be sufficient to balance the internal pressure: the membrane ruptures and the DNA is released. The other mechanism involves a change in the properties of the membrane itself. Part of the membrane is composed of a polyelectrolyte bilayer formed by the cationic block and DNA. This suggests that the permeability of the membrane can be controlled by ionic strength through screening of the electrostatic interactions in the polyelectrolyte bilayer.

The stability of the vesicles against osmotic pressure exerted by the supporting medium is illustrated in Figure 8. Fluorescence (left panels) and phase contrast (right panels) micrographs of the same specimens clearly show the release of DNA once the PEG concentration drops below 5 wt %. The empty vesicles are still visible, but the shape of the membrane has become quite irregular. This irregularity suggests that the membrane has been ruptured due to the excess pressure exerted by the encapsulated DNA. Even after release of the DNA, the fluorescence micrographs show the incorporation of a residual amount of DNA in the membrane. The results are summarized in the stability diagram in Figure 9. Here, the minimum, average, and maximum observed radius of the vesicles is displayed versus the osmotic pressure of the supporting medium. The pressure has been derived from the experimental PEG weight fraction and the empirical relation

$$\log\left(\pi_{PEG}\right) = 4.42 + 0.74\left(wt\%\right)^{0.4} \tag{3}$$

($M_w = 3000$).[25] It is clear that the larger vesicles are less stable; they eject their encapsulated material at higher external pressure (see also Figure 8). In order to confine DNA inside the vesicles, the minimum osmotic pressure exerted by the supporting medium is around $10^5$ N/m$^2$. Notice that the latter value corresponds with the osmotic pressure of an assembly of long DNA molecules with concentration 150 g of DNA/l.[26]

We have also investigated the stability of the vesicles against ionic strength. So far, we have not added any low molecular weight salt; all ions come from the DNA (sodium) and the cationic block (iodide). Figure 10 displays fluorescence micrographs of the vesicles in 43 wt % aqueous PEG solution, but with various amounts of added NaCl. If the salt concentration does not exceed 0.1 M, the vesicles are stable and there is no change in confinement of the encapsulated material. For the highest employed 1 M ionic strength, the DNA is released into the medium. As can be seen in the fluorescence micrograph in Figure 10c, the shape of the empty vesicles remains quite regular and some DNA is still incorporated in the membrane. Contrary to the osmotic pressure induced release mechanism, there is no difference in permeability for DNA between vesicles of different size. These results suggest that the salt influences the permeability of the membrane rather that the membrane ruptures at certain salinity. A plausible explanation is that the membrane is semi-permeable for small ions, which allows Donnan salt partitioning between the aqueous supporting medium and the vesicle. The salt screens the electrostatic interactions between DNA and the cationic block in the polyelectrolyte bilayer. Due to this screening, the bilayer becomes less stable and eventually the membrane becomes permeable for the encapsulated material. Release of DNA, either by salt screening or osmotic pressure, occurs typically within tens of seconds to minutes.

**Clone Vector DNA and the 'Charge Inverse' System.** To demonstrate that there are no restrictions to the molecular weight of the DNA, we have encapsulated pUC18 plasmid (2686 base pairs) in cationic PBd-*b*-P4VPQI vesicles by the same procedure. The polarized light micrograph in Figure 11a shows birefringent, liquid-crystalline capsules after the evaporation of the volatile organic solvent in air. For macroscopic, bulk solutions, a first order phase transition to a cholesteric liquid crystal has been observed with a critical boundary pertaining to the complete disappearance of the isotropic phase at 15 g of DNA/l.[27] Accordingly, since the DNA concentration in the stock solution is 1.4 g/l, the observation of the birefringence indicates that the plasmid is compacted by at least a factor of 10. Sub-micron size, pUC18 capsules can be prepared by sonication of the emulsion prior to the solvent quality induced collapse of the copolymer layer. The size distribution of the capsules after evaporation of the volatile solvent was investigated with scanning electron microscopy. With a sonication power of 25 W applied for 5 minutes, a lognormal distribution was obtained with average capsule diameter 0.7±0.2 μm. The minimum observed diameter was 0.3 µm. As in the case of DNA fragments, the pUC18 capsules can be transferred into aqueous medium to produce vesicles with the help of an osmotic agent. With gel electrophoresis, we have checked that the integrity of the plasmid is preserved after subsequent release induced by either an increase in ionic strength or a decrease in external osmotic pressure. The position of the electrophoresis band and the absence of band smear show that the released plasmid is free and not complexed with cationic copolymer.

Finally, we show that our method is not unique for DNA, but that it can also be used to encapsulate other macromolecules. As an example, we have prepared the charge inverse system: capsules of cationic poly(ethylene imine) (PEI) encapsulated by the anionic diblock poly(styrene-*b*-acrylic acid) (PS-*b*-PA) copolymer. For this purpose, stock solutions of PEI in water and 'crew cut' PS-*b*-PA were mixed in volume ratio 1:9 to form an emulsion. Subsequently, the hydrophobic PS attachment was collapsed and dry capsules were made according to the procedure as described above for the encapsulation of DNA. A typical light micrograph of these capsules is displayed in Figure 11b. Vesicles can subsequently be produced by transferring them into water, provided they are stabilized with PEG.

## Conclusions

We have demonstrated that we can achieve efficient encapsulation and at least a 10-fold compaction of short fragment DNA with a cationic diblock copolymer and a single emulsion technique. In principle, there are no restrictions to the molecular weight of the DNA and there is no need for precipitation onto template particles, since the material is inserted in the emulsion droplets before the membrane has been formed. Another advantage is that the size of the vesicles can be controlled by the emulsification procedure. To illustrate these features, we also have prepared sub-micron size vesicles loaded with clone vector, pUC18 plasmid. Around 5 percent of the vesicles with short fragment DNA exhibit a cholesteric texture under crossed polarizers. The remaining fraction does not show a periodicity in light intensity in the radial direction, which suggests that the molecules are hexagonally ordered. They do not show, however, the characteristic fan-like shapes. This might be related to the mesoscopic dimension and the geometric frustration of the spherical copolymer interface. The DNA inside the vesicles is tightly packed. As derived from the concentration, the interaxial spacing between the molecules is around 3.3 nm, which is similar to that in DNA condensates and phage heads.[4,5]

The mechanism for the confinement of the DNA inside the vesicles is the elasticity of the membrane formed by the complex of the collapsed copolymer and DNA. We have achieved long-time stability of the vesicles by using an osmotic agent in the supporting medium. Apart from stability, controlled release is also of great importance for practical applications. We have demonstrated that the DNA is released once the osmotic pressure of the supporting medium drops below a certain critical value. The larger vesicles are less stable against osmotic pressure, which indicates release by rupture of the membrane. Another characteristic property is the sensitivity to ionic strength. The stability and permeability of the membrane depend on the electrostatic interactions between the ionic block of the copolymer and DNA. At very high salinity (around 1 M), the electrostatic interactions in the polyelectrolyte bilayer are effectively screened and the membrane becomes permeable for the encapsulated material. It was checked with electrophoresis that the released plasmid is truly free and not complexed with copolymer. This indicates the absence of co-adsorbed cationic copolymer inside the vesicles. In this respect, our method differs from the classical preparation of non-viral gene delivery systems, where *all* DNA is complexed with cationic polymer.

We have demonstrated that the encapsulation method can also be used to encapsulate other charged (bio)polymers. As an illustrative example, we have prepared the inverse system: capsules of cationic poly(ethylene imine) encapsulated by the anionic diblock poly(styrene-*b*-acrylic acid) copolymer. Since the copolymer is also the building block of the membrane, its chemical composition and molecular weight control the stability and functionality of the vesicle. In particular, the stability of the self-assembled structure in various environments needs to be optimized to suit the particular application. A promising option is chemical cross-linking (polymerizing) of the collapsed polymer layer. Other promising features are the possibilities to control biodegradability and tissue-specific adaptation by the specific choice of copolymer.

**Acknowledgement.** We are grateful to C. Woldringh for use of his laboratory for the fluorescence microscopy experiments. We thank S. Cunha and A. Kros for assistance in fluorescence and scanning electron microscopy, respectively.

**Table 1.** Slopes $\alpha$ resulting from the fit of eq (2) to the fluorescence intensities in Figure 7 for emulsion droplets in toluene, capsules in air, and vesicles in aqueous PEG solution. Notice that the margins refer to a variation in DNA concentration between capsules or vesicles of comparable size, rather than uncertainty in the measurements.

| | $\alpha$ (au) | $C_{DNA}$ (g/l) |
|---|---|---|
| Droplets | 0.08±0.01 | 36 |
| Capsules | 0.5±0.1 | 240±40 |
| Vesicles | 0.8±0.1 | 350±50 |

## Legends to the figures

**Figure 1**        Schematic representation of an emulsion droplet in a good (top) and poor (bottom) solvent for the hydrophobic chain part. The change of the solvent quality results in a collapse of the interfacial polymer layer.

**Figure 2**        Light microscopy image of a DNA emulsion stabilized with PBd-*b*-P4VPQ diblock copolymer and suspended in toluene. The DNA concentration inside the droplets amounts 36 g/l and the initial copolymer concentration in toluene is 4 g/l.

**Figure 3**        Polarized light microscopy of the encapsulation process: (a) emulsion phase in a mixture of toluene and ethyl acetate; (b) capsules after evaporation of the volatile organic solvent in air; (c) vesicles immersed in aqueous PEG solution ($C_{PEG}$ = 43 wt %); (d) as in (c) but for cholesteric vesicles. Notice that the micrograph in (a) shows the coexistence of emulsion droplets and birefringent DNA capsules.

**Figure 4**        Scanning electron microscopy images of DNA capsules at two different magnifications indicated by the bars.

**Figure 5**        Fluorescence micrographs showing labeling of DNA by DAPI; (a) emulsion droplets in toluene; (b) capsules in air; (c) vesicles immersed in aqueous PEG solution ($C_{PEG}$ = 43 wt %).

**Figure 6**        Radial distribution of the fluorescence intensity of DAPI labeled DNA inside an emulsion droplet in toluene (squares), a capsule in air (circles), and a vesicle immersed in aqueous PEG solution (43 wt %, triangles). The intensities are normalized to the intensity at the center of the 2D image of the object, whereas the radial coordinate is scaled by the radius $R$. The solid curve represents a fit of a uniform profile eq (1) with $R$ = 1.5, 2.6, and 4.8 μm in toluene, air, and PEG solution, respectively.

**Figure 7**     Fluorescence intensity at the center of the 2D image of droplets in toluene (triangles), capsules in air (circles), and vesicles in aqueous PEG solution (0.43 wt %, squares). The solid lines represent linear fits of eq (2) with slopes collected in Table 1. Notice that the scattering of the data is largely due to the variation in DNA concentration inside capsules or vesicles of comparable size.

**Figure 8**     Fluorescence (left panels) and phase contrast (right panels) micrographs showing the stabilization of the vesicles by osmotic stress exerted by PEG in aqueous solution. From top to bottom the PEG concentration decreases according to $C_{PEG}$ = 43 (a), 21 (b), 10 (c), and 1 wt % (d). Notice that DNA is released for $C_{PEG}$ = 1 wt %.

**Figure 9**     Minimum (circles), average (triangles), and maximum (squares) radius of the vesicles versus the osmotic pressure exerted by PEG in aqueous solution.

**Figure 10**     Fluorescence micrographs showing the destabilizing of the capsules by an increase of the ionic strength of the supporting medium: (a) no added salt, (b) 0.1 M NaCl, and (c) 1 M NaCl. The PEG concentration amounts $C_{PEG}$ = 43 wt %. Notice that for salt concentrations exceeding 0.1 M the DNA is released.

**Figure 11**     Light micrographs of (a) pUC18 plasmid encapsulated by cationic PBd-*b*-P4VPQ (between crossed polarizers) and (b) PEI encapsulated by anionic PS-*b*-PA.

# Figure 1

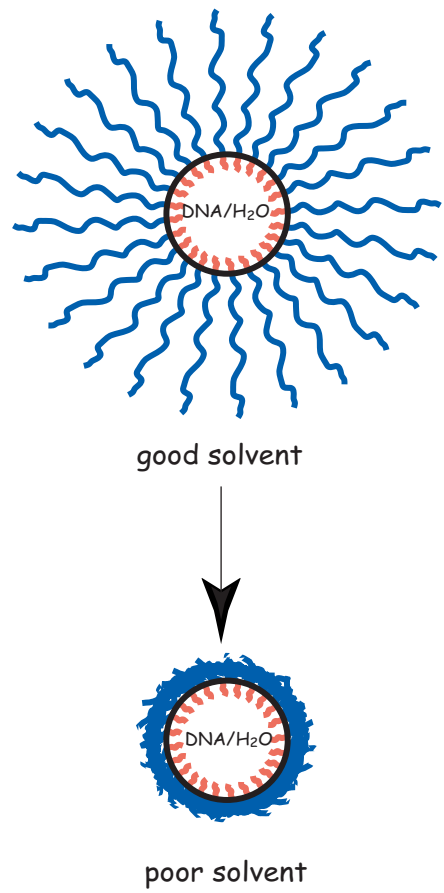

good solvent

poor solvent

# Figure 2

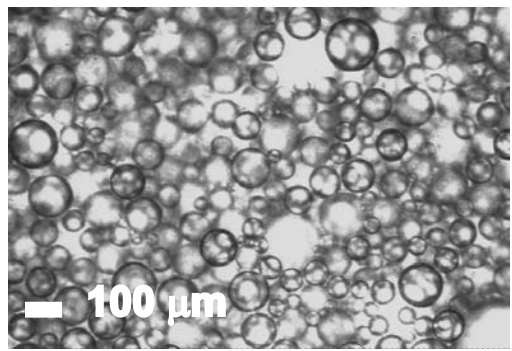

# Figure 3

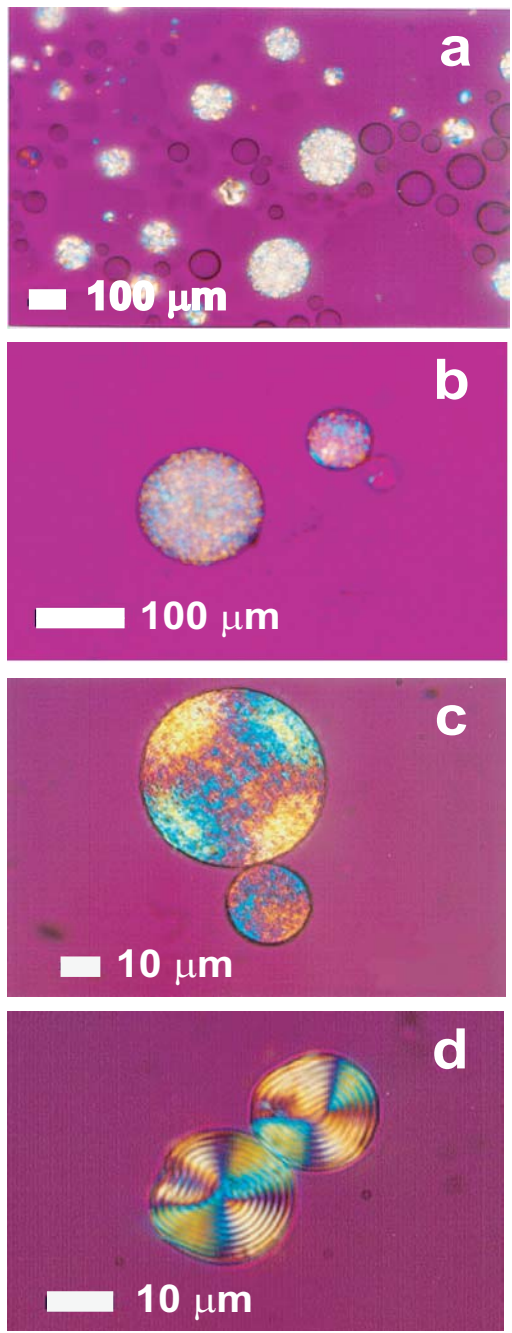

# Figure 4

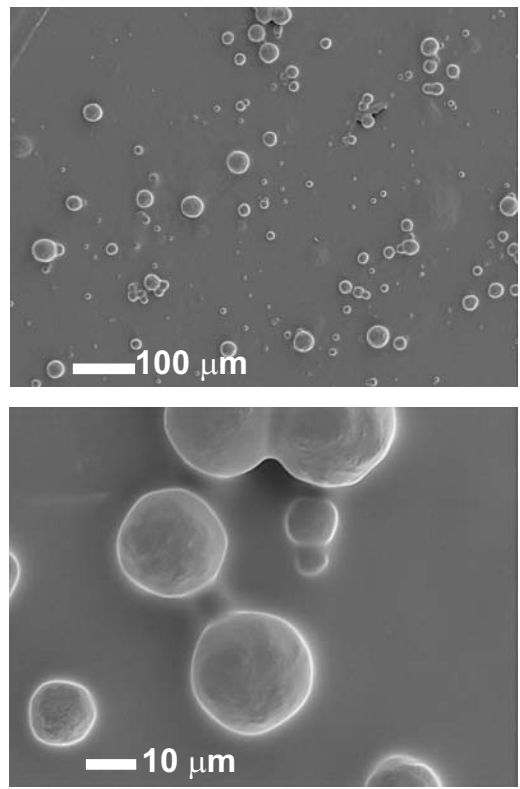

# Figure 5

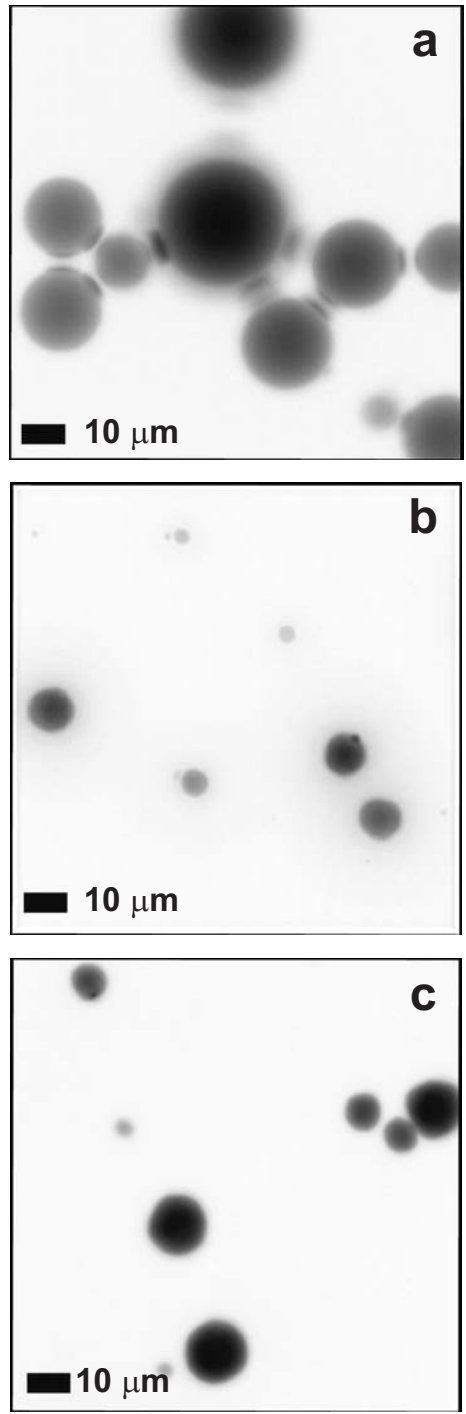

**Figure 6**

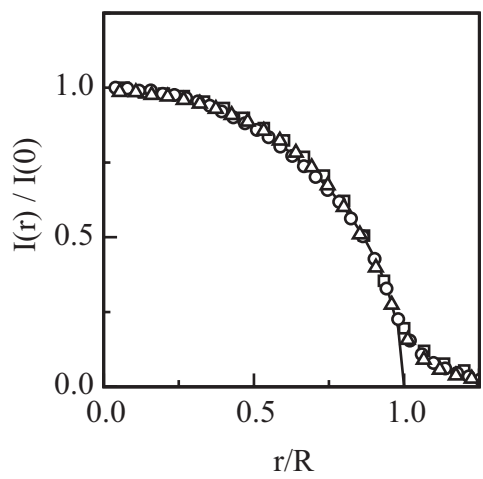

**Figure 7**

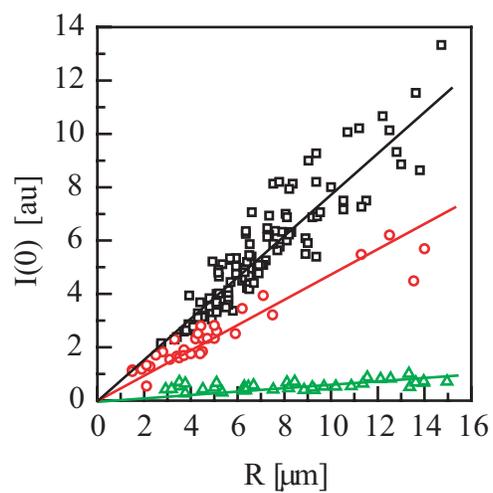

# Figure 8

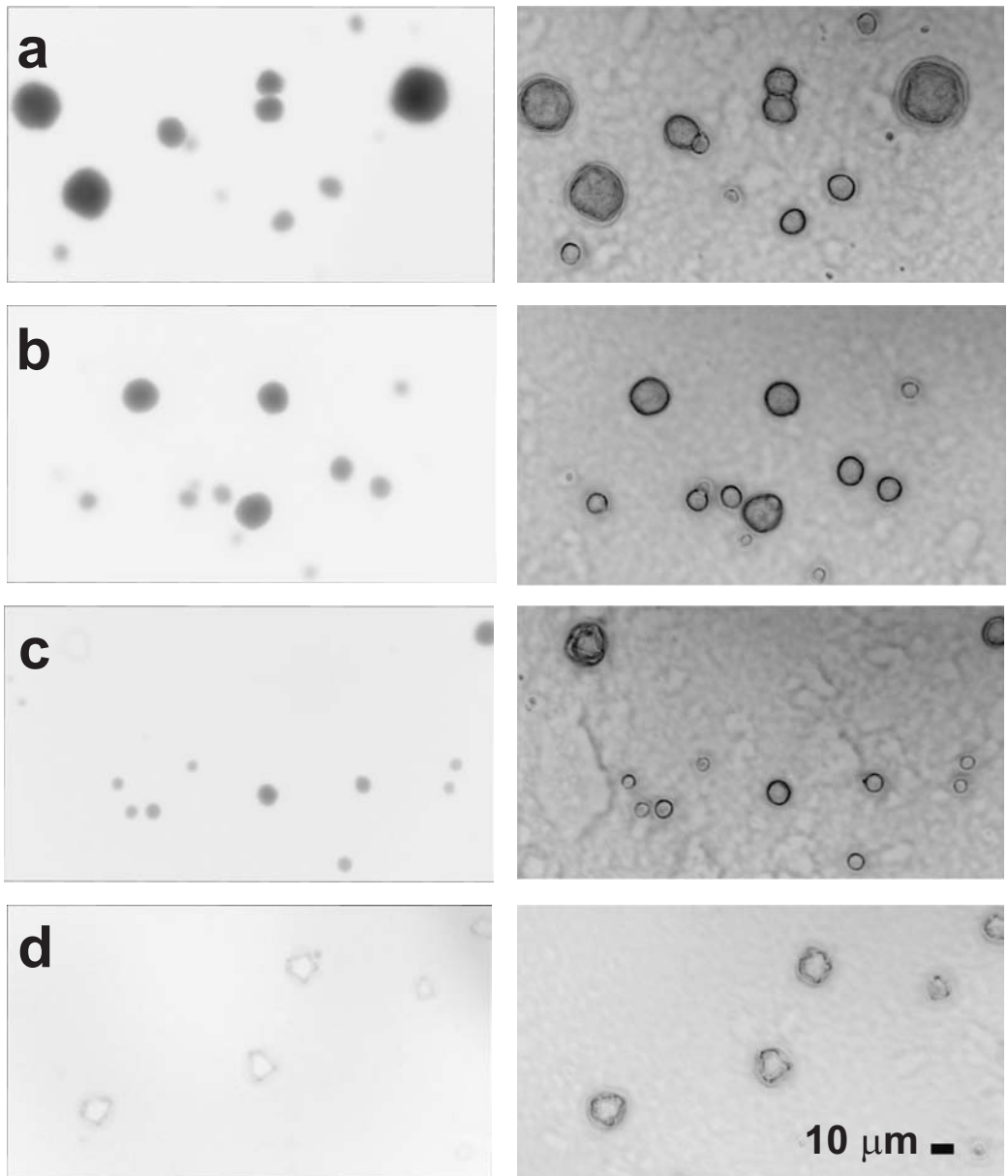

**Figure 9**

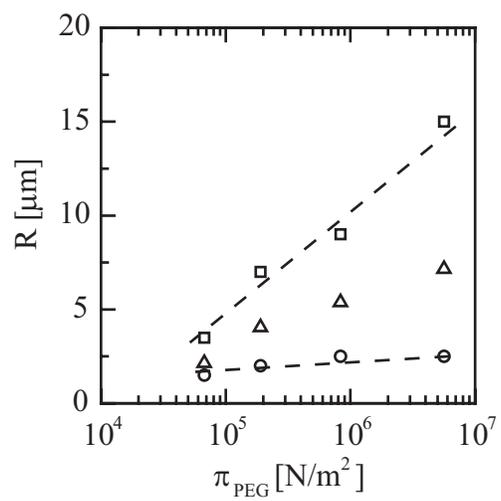

# Figure 10

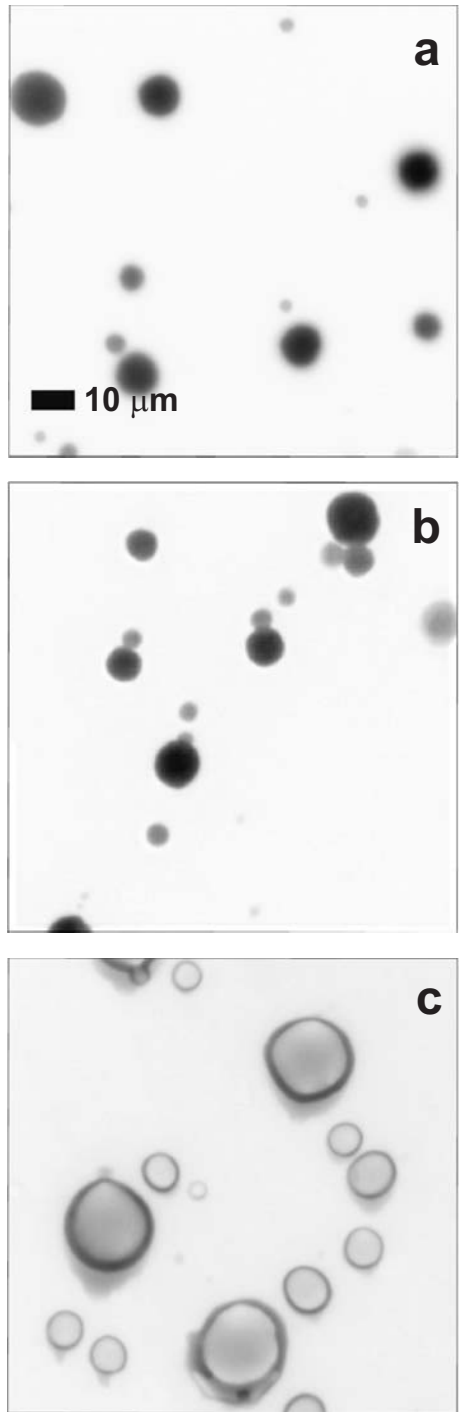

**Figure 11**

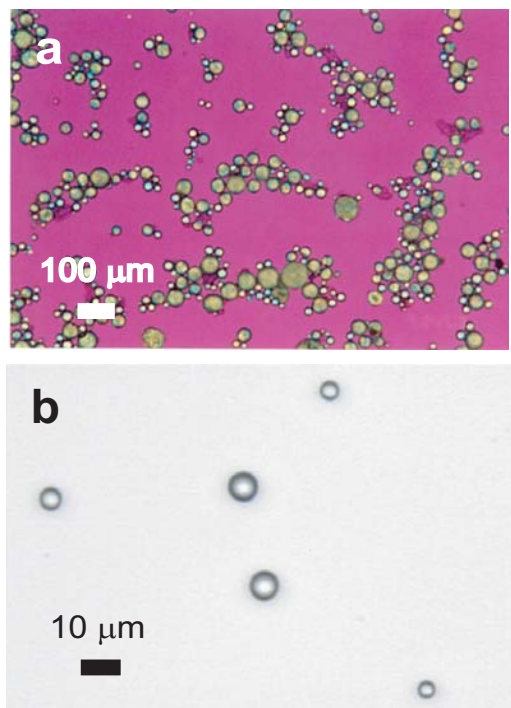

# References

(1)    Zasadzinski, J. A.; Kisak, E.; Evans, C. *Curr. Opin. Coll. & Interf. Sci.* **2001**, *6*, 85-90.

(2)    Discher, B. M.; Hammer, D. A.; Bates, F. S.; Discher, D. E. *Curr. Opin. Coll. & Interf. Sci.* **2000**, *5*, 125-131.

(3)    Vijayanathan, V.; Thomas, T.; Thomas, T. J. *Biochemistry* **2002**, *41*, 14085-14094.

(4)    Bloomfield, V. A, *Biopolymers* **1997,** *44*, 269-282.

(5)    Cerritelli, M. E.; Cheng, N. Q.; Rosenberg, A. H.; McPherson, C. E.; Booy, F. P.; Steven, A. C. *Cell* **1997**, 91, 271-280.

(6)    Odijk, T.; Slok, F. *J. Phys. Chem.* **2003**, 107, 8074-8077.

(7)    Decher, G. *Science* **1997**, *277*, 1232-1237.

(8)    Moehwald, H. *Colloids Surf A* **2000**, *171*, 25-31.

(9)    Shchukin, D. G.; Patel, A. A.; Sukhorukov, G. B.; Lvov, Y. M. *J. Am. Chem. Soc.* **2004**, *126*, 3374-3375.

(10)   Arigita, C.; Zuidam, N. J.; Crommelin, D. J. A.; Hennink, W. E. *Pharm. Res.* **1999**, *16*, 1534-1541.

(11)   Kakizawa, Y.; Kataoka, K. *Adv. Drug Deliv. Rev.* **2002**, *54*, 203-222.

(12)   Reschel, T.; Konak, C.; Oupicky, D.; Seymour, L.W.; Ullbrich, K. *J. Control. Release* **2002**, *81*, 201-217.

(13)   Bruinsma, R. Eur. Phys. *J. B* **1998**, *4*, 75-88.

(14)   Gebhart, C. L.; Kabanov, A. V. *J. Control. Release* **2001**, *73*, 401-416.

(15)   Pautot, S.; Frisken, B. J.; Weitz, D. A. *Langmuir* **2003**, *19*, 2870-2879.

(16)   Wang, L.; Ferrari, M.; Bloomfield, V. A. *BioTechniques* **1990**, *9*, 24-27.

(17)    Nicolai, T.; van Dijk, L.; van Dijk, J. A. P. P.; Smit, J. A. M. *J. Chromatogr.* **1987**, *389*, 286-292.

(18)    Liebe, D. C.; Stuehr, J. E. *Biopolymers* **1972**, *11*, 167-184.

(19)    Zakharova, S. S.; Jesse, W.; Backendorf, C.; van der Maarel, J. R. C. *Biophys. J.* **2002**, *83*, 1106-1118.

(20)    http://simon.bio.uva.nl/object-image.html

(21)    Kassapidou, K.; Jesse, W.; van Dijk, J. A. P. P.; van der Maarel, J.R.C. *Biopolymers* **1998**, *46*, 31-37.

(22)    Rill, R. L.; Strzelecka, T. E.; Davidson, M. W.; Van Winkle, D. H. *Physica A* **1991**, *176*, 87-116.

(23)    Livolant, F.; Leforestier, A. *Prog. Polym. Sci.* **1996**, *21*, 1115-1164.

(24)    Kassapidou, K.; van der Maarel, J. R. C. *Eur. Phys.  J. B* **1998**, *3*, 471-476.

(25)    Reid, C.; Rand, R. P. *Biophys. J.* **1997**, 72, 1022-1030.
       http://aqueous.labs.brocku.ca/osfile.html

(26)    Strey, H. H.; Parsegian, V. A.; Podgornik, R. *Phys. Rev. Letters* **1997**, *78*, 895-898.

(27)    Zakharova, S. S.; Jesse, W.; Backendorf, C.; van der Maarel, J. R. C. *Biophys. J.* **2002**, *83*, 1119-1129.